*Article*

# Fe$_{1-x}$Ni$_x$ Alloy Nanoparticles Encapsulated inside Carbon Nanotubes: Controlled Synthesis, Structure and Magnetic Properties

**Rasha Ghunaim [1,2,\*], Christine Damm [1], Daniel Wolf [1], Axel Lubk [1], Bernd Büchner [1,3], Michael Mertig [2,4] and Silke Hampel [1]**

[1] Leibniz Institute for Solid State and Material Research Dresden, Helmholtzstrasse. 20, 01069 Dresden, Germany; C.Damm@ifw-dresden.de (C.D.); d.wolf@ifw-dresden.de (D.W.); a.lubk@ifw-dresden.de (A.L.); B.Buechner@ifw-dresden.de (B.B.); S.Hampel@ifw-dresden.de (S.H.)
[2] Institute for Physical Chemistry, Technische Universitaet Dresden, 01062 Dresden, Germany; michael.mertig@tu-dresden.de
[3] Institute for Solid State Physics, Technische Universitaet Dresden, 01062 Dresden, Germany
[4] Kurt-Schwabe-Institut für Mess- und Sensortechnik e. V. Meinsberg, 04736 Waldheim, Germany
\* Correspondence: r.ghunaim@ifw-dresden.de (R.G.); Tel.: +49-(0)351-4659-413

**Abstract:** In the present work, different synthesis procedures have been demonstrated to fill carbon nanotubes (CNTs) with Fe$_{1-x}$Ni$_x$ alloy nanoparticles (x = 0.33, 0.5). CNTs act as templates for the encapsulation of magnetic nanoparticles, and provide a protective shield against oxidation as well as prevent nanoparticles agglomeration. By variation of the reaction parameters, the purity of the samples, degree of filling, the composition and size of filling nanoparticles have been tailored and therefore the magnetic properties. The samples were analyzed by scanning electron microscopy (SEM), transmission electron microscopy (TEM), Bright-field (BF) TEM tomography, X-ray powder diffraction, superconducting quantum interference device (SQUID) and thermogravimetric analysis (TGA). The Fe$_{1-x}$Ni$_x$-filled CNTs show a huge enhancement in the coercive fields compared to the corresponding bulk materials, which make them excellent candidates for several applications such as magnetic storage devices.

## 1. Introduction

The synthesis and characterization of nanomaterials are very exciting and emerging research fields which have received much attention in the scientific and technological fields [1]. The possibility to fabricate nanomaterials and nanocomposites opens the door to numerous applications in multidiscipline fields, such as magnetic storage devices [2,3], fuel cells [4], electromagnetic waves absorption [5], sensors for magnetic force microscopy [6] and human tumor therapy [7–9]. FeNi binary alloys exhibit special properties, such as shape memory, soft magnetization in which their saturation magnetization increases with increasing iron content, and martensitic bcc-fcc phase transition with increasing nickel concentration. Due to their unique properties, FeNi alloys can be used in air craft engines, steam turbines in power generation plants, medicine, and in nuclear, chemical and photochemical industries [10,11].

Due to the enhancement of materials properties, when their size scaled down to the nano-regime, many studies have reported on the synthesis and properties of FeNi binary nanoparticles and nanowires. McNerny et al. reported the synthesis of a series of monodisperse FeNi magnetic nanoparticles (MNPs) with tunable Curie temperature ($T_c$) for self-regulated hyperthermia [12]. N-doped carbon nanotubes filled with FeNi alloy nanoparticles can be used as counter electrode in dye-sensitized solar cells (DSSCs) in comparison to Pt-counter electrode. This is due to the modification of the surface electronic properties of carbon nanotubes by the encapsulated FeNi nanoparticles [13]. Theoretical calculations showed that permalloy nanowires-filled CNTs, or nickel-rich FeNi alloys [14], are attractive candidates for the new types of electromagnetic waves absorber with thinner coating thickness [15]. On the other hand, iron-rich FeNi alloys, called Invar $Fe_{65}Ni_{35}$, discovered by Guillaume in 1897, are of particular interest [16]. These alloys have extremely low thermal expansion coefficient (ca. one tenth of that of steel), making them excellent candidates for the fabrication of electronic devices in aircraft controls, laser systems and bimetallic thermostats [17].

Due to their high surface-to-volume (S/V) ratio, MNPs are more susceptible to harsh environments such as oxidation, agglomeration and aggregation [18,19]. Therefore, it is necessary to produce MNPs with protective layers which preserve their properties. CNTs have been introduced as protective shells due to their high stability in different chemical and physical environments such as acids, bases, high temperature and pressure media [20–23]. CNTs are also able to control the size and morphology of the filling material due to the confinement of the MNPs within the hollow tubular cavity of CNTs. Chemical vapor deposition (CVD) is a technique used to fill CNTs with MNPs as in-situ filling, in which metallocenes precursors are used as a source for carbon and MNPs [24]. Hydrocarbons (such as benzene) can also be used as carbon precursors, which can be decomposed in an inert atmosphere over freshly prepared FeNi alloys [20]. Permalloy-filled CNTs [15] or Invar alloy-filled CNTs [17] have also been prepared via CVD. However, post-synthesis filling is a facile method for the filling of CNTs via wet chemistry. CNTs can work as molecular straws which are able to absorb materials via capillary forces [25–27]. Wu et al. reported the filling of CNTs with FeNi nanoparticles alloys consisting of 75 at. % Fe and 25 at. % Ni via wet chemistry, using salts nitrates as metals precursors [28].

Due to the attractive properties of FeNi alloys, this study was directed towards the synthesis of CNTs-based nanocomposites of FeNi magnetic nanoparticles based on post-synthesis method, in which pre-fabricated multi-walled CNTs (MWCNTs) were used as templates. Two facile filling approaches were applied for the filling of CNTs with two different stoichiometries of $Fe_{1-x}Ni_x$ alloy nanoparticles (x = 0.33, 0.5). A study on the morphology, structure and magnetic properties for the resulting FeNi magnetic nanoparticles is presented. The promising effect of an additional heat-treatment step for $Fe_{1-x}Ni_x$ alloy nanoparticles (x = 0.33) on these properties was investigated.

## 2. Materials and Methods

### 2.1. Preparation of the Binary Alloys Inside CNTs

Multi-walled carbon nanotubes (MWCNTs) of the type PR-24-XT-HHT (Pyrograf products, Inc., Cedarville OH, USA) were used as templates or nano-containers for the preparation of intermetallic nanoparticles. This type of CNTs is distinguished by its high purity which results from the fact that the as-produced carbon nanotubes are heat treated to 3000 °C. This procedure reduces the iron content (i.e., the catalyst) to a very low level (< 100 ppm) [29–31]. Previously, we reported the filling of CNTs with FeCo binary alloys by following two approaches of filling [32]. In this study, we adapted these approaches for the preparation of $Fe_{50}Ni_{50}$@CNT and $Fe_{67}Ni_{33}$@CNT samples. They are briefly described as follows.

The first approach is an extension of a reported solution filling approach for CNTs [33]. Standard aqueous solutions (1 M) of $Fe(NO_3)_3 \cdot 9H_2O$ (grade: ACS 99.0–100.2%) and $Ni(NO_3)_2 \cdot 6H_2O$ (ACS 99.0–102.0%) were prepared and combined in a stoichiometric ratio with respect to the metal ions (i.e., Fe:Ni = 1:1 or 2:1). MWNTs (50–100 mg) were added and the mixture was treated in an ultrasonic bath for 1 h (for $Fe_{50}Ni_{50}$) and 45 min (for $Fe_{67}Ni_{33}$) at room temperature. The mixture was then vacuum-filtered and washed with acetone and distilled water (20–30 mL) of a volumetric ratio of 1:1. The solid residue was then dried for 24 h at a temperature of 100 °C and reduced under hydrogen and argon atmosphere (50 vol. % $H_2$ + 50 vol. % Ar) at a temperature of 500 °C for 4 h. An additional heat treatment step was essential for the $Fe_{67}Ni_{33}$@CNT samples to obtain the desired intermetallic phase, in which the reduced samples were further annealed under a mixture of Ar and $H_2$ gases streams (95 vol. % Ar + 5 vol. % $H_2$) at a temperature of 500 °C for 12 h [21,34,35].

In the second approach of filling, the nitrate precursors were directly mixed with the specific amount of CNTs (50–100 mg) in a sealed round bottom flask. A few drops of distilled water were added to ensure good stirring. The mixture was placed in an oil bath and heated to a temperature of around T ≈ 65 °C for 4 h. The mixture was then naturally cooled down to room temperature and washed with acetone and distilled water (20–30 mL) of a volumetric ratio of 1:1. The samples were then dried, reduced and annealed in a manner similar to the solution filled samples. This approach differs from the solution one, in the way that the filling materials are provided with a higher concentration (due to the absence of water solvent), which increases the percentage of materials which fill the CNTs.

*2.2. Characterization*

All samples were routinely investigated by scanning electron microscopy (SEM) with a Nova 200 NanoSEM from FEI Company operated at 15 kV and combined with energy dispersive X-ray (EDX) analyzer (AMETEK). The SEM samples were prepared by placing a thin film of the sample on a carbon tape. Transmission electron microscopy (TEM), high-resolution transmission electron microscopy (HRTEM) measurements and nanobeam electron diffraction patterns were performed using a Tecnai F30 (FEI) operated at 300 kV or a Tecnai G2 (FEI) operated at 200 kV. Both are equipped with EDX analyzer (AMETEK, Oxford). Bright-field (BF) TEM tomography was employed to reveal the 3D positions of the filling FeNi nanoparticles and the morphology of the CNTs. To this end, a tilt series of BFTEM micrographs within a tilt range from -70° to +62° and tilt steps of 3° were recorded and reconstructed. The TEM samples were prepared by adding a few droplets of the sample suspension in acetone on a copper grid with a carbon coating on one side.

Crystal structure determinations for the magnetic nanoparticles inside CNTs were identified by X`Pert Pro MPD PW3040/60 X-ray diffractometer (XRD) (PANALYTICAL) with Co K$\alpha$ radiation ($\lambda$ = 1.79278 Å) in reflection geometry at a scanning rate of 0.05° $s^{-1}$ in the 2$\theta$ range from 10° to 80°.

Thermogravimetric measurements (TGA) were performed with a SDT-Q600 (TA instruments). A few milligrams of the material (~5 mg) were heated to a temperature of 900 °C with a heating rate of 5 K/min followed by an isothermal of 15 min under air atmosphere with a flow rate of 100 mL/min.

The magnetic field dependence of the magnetization at 5 K and 300 K for an external magnetic field up to ± 3 T was measured by means of superconducting quantum interference device (MPMS-XL SQUID) magnetometer from Quantum Design (San Diego, CA, USA). Samples were filled inside gelatin capsules.

## 3. Results and Discussions

This section is divided into two parts: the first part summarizes all the data obtained for $Fe_{50}Ni_{50}$@CNT samples, while the second part includes all $Fe_{67}Ni_{33}$@CNT samples data.

*3.1. .$Fe_{50}Ni_{50}$@CNT*

3.1.1. Morphology and Structure

The morphology and geometry of the filling material and their location inside or outside CNTs were examined by SEM (Figure 1).

Figure 1a shows an overview image in back scattered electron (BSE) mode for $Fe_{50}Ni_{50}$@CNT samples prepared by the first (i.e., solution) filling approach. The filling particles are distributed along the inner cavity of the hollow CNTs. For samples prepared by the second filling approach (Figure 1b), the same behavior is observed, however with a seemingly higher degree of filling compared to the solution approach. This can be seen from the pearl necklace-like appearance of the filling inside the CNTs and was confirmed by quantitative measurements performed by TGA as will be shown later.

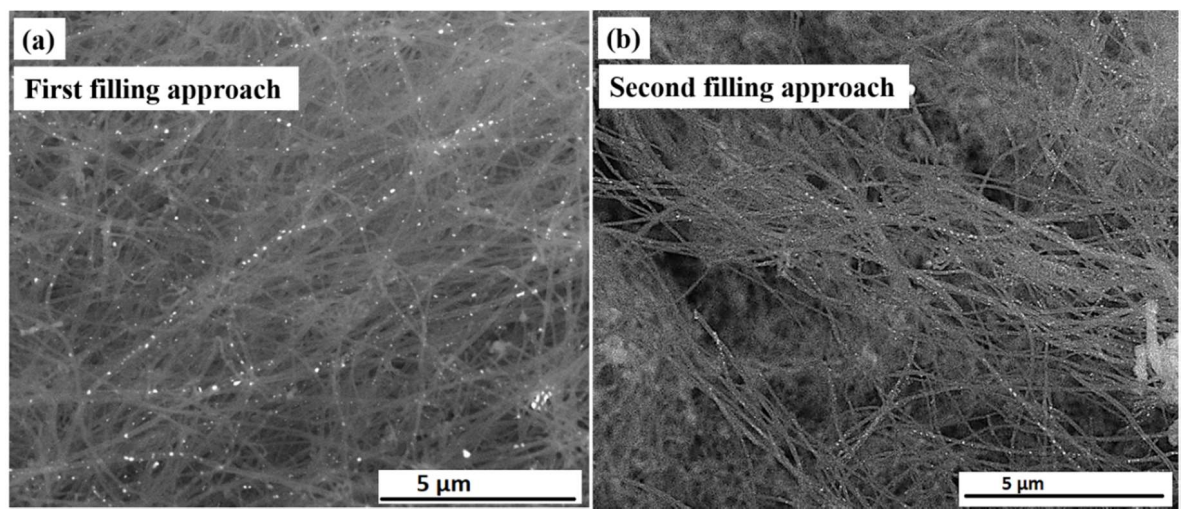

**Figure 1.** SEM overview images in BSE contrast for Fe$_{50}$Ni$_{50}$@CNT samples prepared by the: (**a**) first filing approach (solution); and (**b**) second filing approach.

TEM measurements were performed for samples prepared by both filling approaches. Figure 2 shows samples of Fe$_{50}$Ni$_{50}$@CNT prepared by the first (Figure 2a) and second (Figure 2d) filling approaches, in which most of the particles are located within the hollow cavity of the CNTs and exhibit diameters either in the range or smaller than the diameter of the CNTs inner walls.

TEM-based nanobeam electron diffraction measurements were carried out for several individual nanoparticles of Fe$_{50}$Ni$_{50}$@CNT prepared by the first (Figure 2b,c) and second (Figure 2e) filling approaches. The measurements revealed that the filling particles are single crystalline, as indicated by the reflections corresponding to the 111 (0.206 nm), 200 (0.176 nm), 220 (0.124 nm), 311 (0.107 nm), 222 (0.104 nm) and 400 (0.088 nm) lattice planes confirming the fcc structure of FeNi phase. A 3D reconstruction of the FeNi- filled CNTs obtained by TEM tomography strongly evidences the location of the filling nanoparticles inside the hollow cavity of the CNTs (Figure 2f).

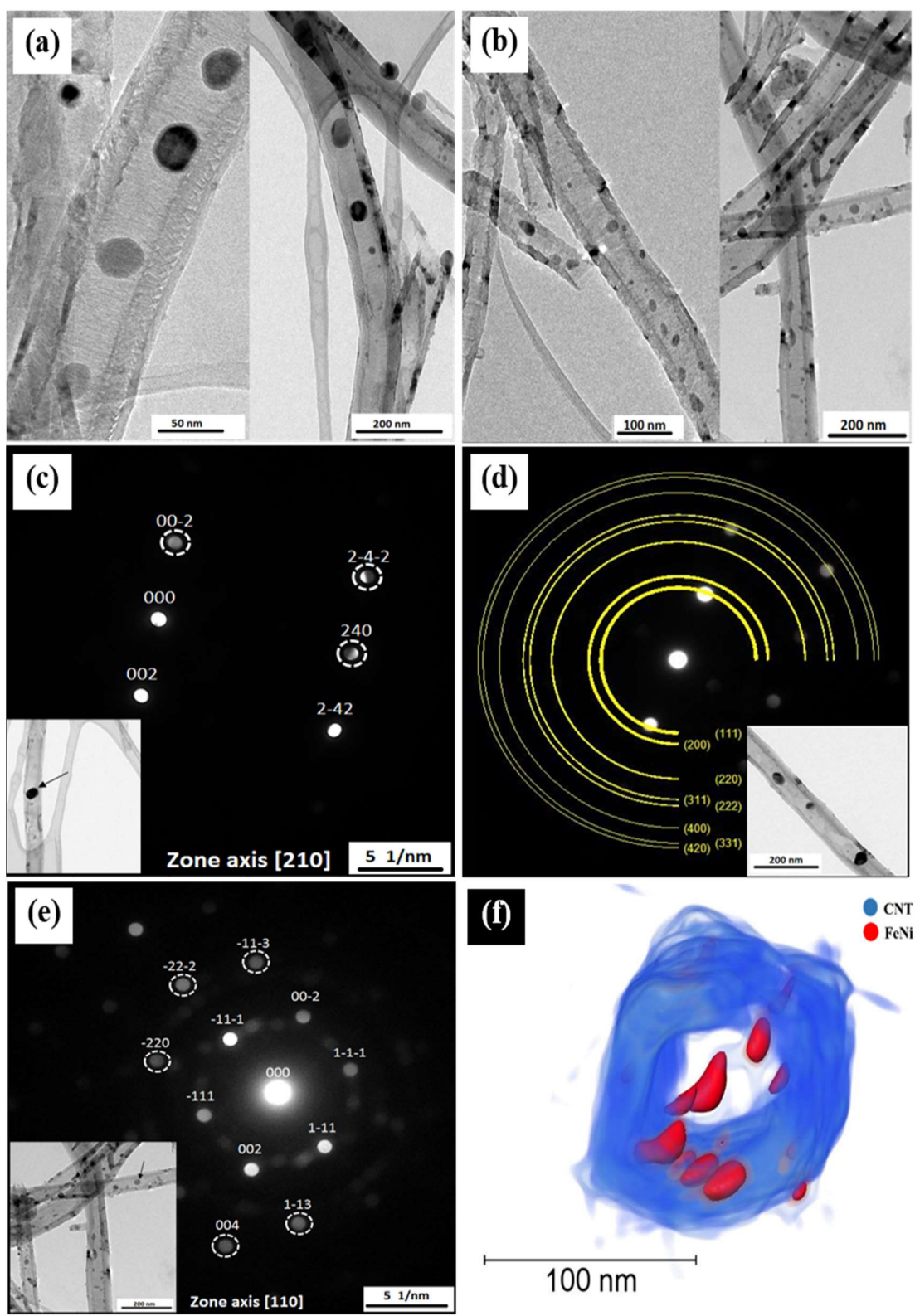

**Figure 2.** TEM bright field images of Fe$_{50}$Ni$_{50}$@CNT samples prepared by the: (**a**) first filling approach (solution); and (**b**) second filling approach. Nanobeam electron diffraction patterns for selected nanoparticles prepared by the (c, d) first and (e) second filling approaches with the corresponding TEM images as insets. (**f**) 3D volume rendering of FeNi nanoparticles inside CNTs reconstructed by TEM tomography.

The distribution of the particles diameter was investigated for the samples prepared by both filling approaches. The diameters were measured perpendicular to the long axis of the CNTs. Fe$_{50}$Ni$_{50}$@CNT nanoparticles prepared by the first filling approach have a mean diameter of d$_{TEM}$ = 19 ± 10 nm (Figure 3a), whereas nanoparticles prepared by the second filling approach, a mean diameter of d$_{TEM}$ = 17 ± 4 nm was found (Figure 3b), values which are comparable with the mean diameter of the hollow cavity of the CNTs (d$_{CNT}$ = 42 ± 14 nm) (Figure 2). It is worth mentioning that the error bars in d$_{TEM}$ refer to the observed size distribution.

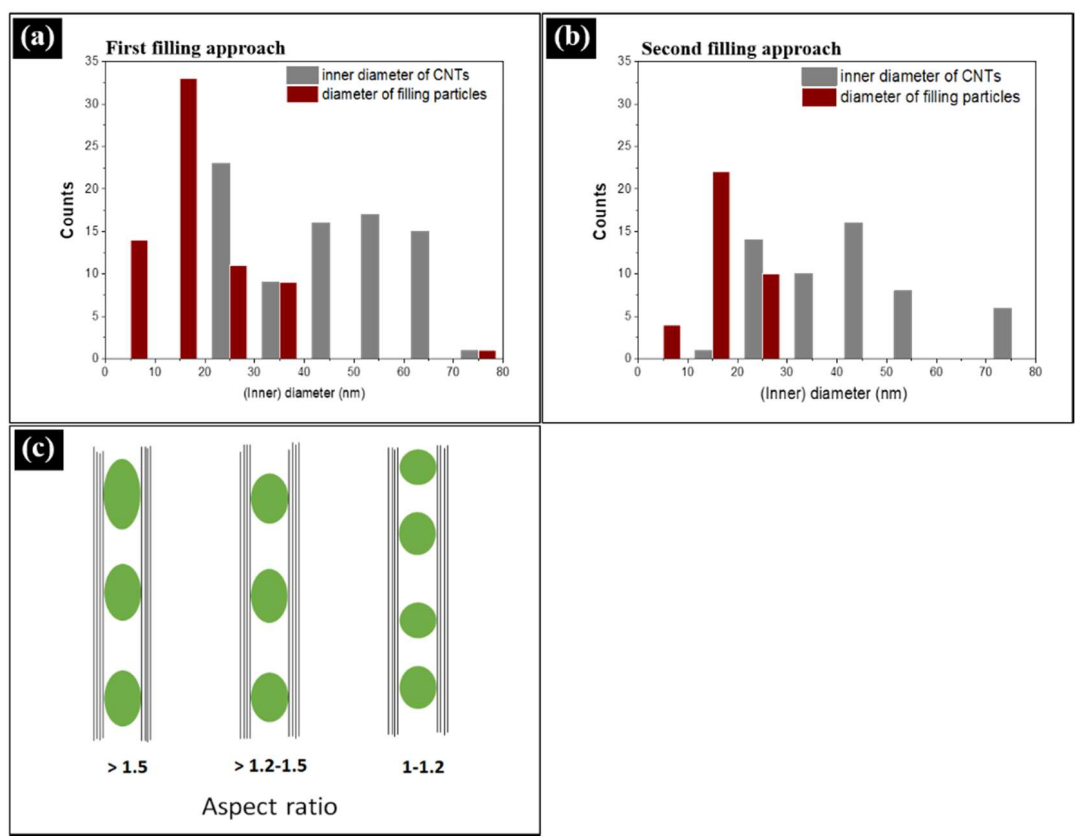

**Figure 3.** Histograms representing the size distribution of the inner diameter (nm) of CNTs and particles diameters for Fe$_{50}$Ni$_{50}$@CNT samples prepared by the: (**a**) first filing approach; and (**b**) second filling approach. (**c**) Schematic representation of the geometry of the filling particles with respect to the aspect ratio values.

The spherical geometry of the intermetallic nanoparticles inside CNTs was confirmed by TEM aspect-ratio studies (i.e., the ratio of the particle's long axis to its short axis).

In a sample prepared by the first filling approach (with nearly 68 investigated filling particles), 81% of the particles had ratios in the range of 1–1.2, whereas 18% of the particles had ratios in the range of 1.2–1.5. The remaining 1% exhibited ratios larger than 1.5. On the other hand, in a sample prepared by the second filling approach (with nearly 36 investigated filling particles), 58% of the particles had aspect ratios in the range of 1–1.2, whereas 22% had ratios in the range of 1.2–1.5. The remaining 20% exhibited ratios larger than 1.5. Hence, the majority of the filling particles have a spherical geometry. Figure 3c shows a schematic representation of the geometry of the filling particles based on the aspect ratio values. It is worth mentioning, that the percentage of particles which exhibited ratios larger than 1.5 was found to be higher in the second approach (20%) compared to the first approach (1%). This higher value gave a hint about the higher degree of filling in the second approach, which allowed a continuous behavior of filling within the tubes. However, in the first approach of filling, a lower degree of filling allowed the filling particles to distribute separately within the tubes cavity. This, in turn, allowed most of the filling particles (81%) to have a uniform spherical diameter with aspect ratio of 1–1.2 compared to 58% in the second filling approach.

The expected stoichiometry of 1:1 for the binary alloys of Fe$_{50}$Ni$_{50}$@CNT was confirmed by EDX measurements in SEM and TEM. In SEM-EDX, the stoichiometry was obtained by measuring the relative ratio of the individual elements over an analyzed area of about 60 × 50 μm² (see Figure S1 in the Supplementary Materials), whereas, in TEM-EDX, the relative ratio was obtained for a large number of individual particles. Quantitative analysis indicates that the average atomic percentage of Fe is 48.0 ± 1.0 at. % and for Ni is 52.0 ± 1.0 at. % in Fe$_{50}$Ni$_{50}$@CNT samples prepared by both filling approaches.

The XRD diffraction pattern for Fe$_{50}$Ni$_{50}$@CNT prepared by both filling approaches is shown in Figure 4. The intense reflection at 2θ ~ 30° corresponds to the 002 lattice plane of CNTs (labeled with C). The reflections at 2θ ~ 51° and 60° correspond to the lattice planes 111 and 200, respectively, of the fcc structure of FeNi with space group Fm-3m (225, cubic, PDF No. 04-004-8844). [24]. No reflections corresponding to oxides or carbides were detected, emphasizing on the fact that CNTs act as protective shells for the MNPs.

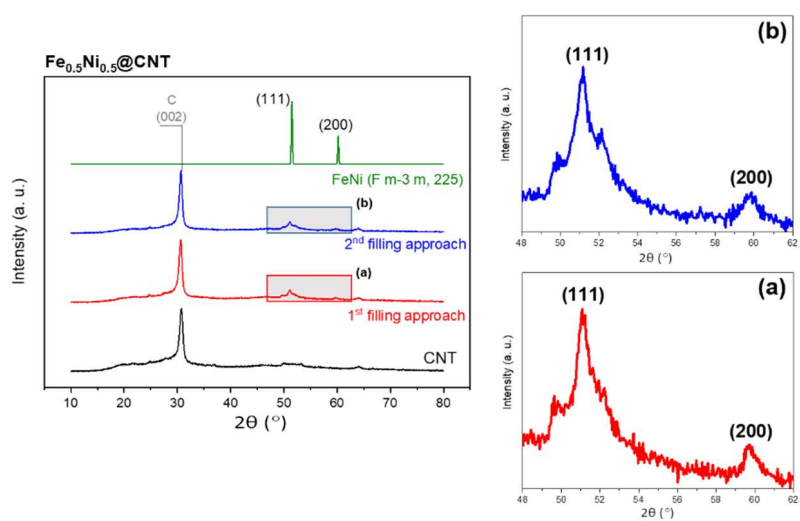

**Figure 4.** XRD diffraction pattern for Fe$_{50}$Ni$_{50}$@CNT sample prepared by the first and second filling approaches (*left-hand side*). Enlarged views (marked by rectangles) for the main reflections for samples prepared by (**a**) 1$^{st}$ and (**b**) 2$^{nd}$ filling approaches, are shown on the right-hand side.

Assuming the spherical shape of the nanoparticles, the mean particle diameter was calculated using Scherrer's equation, given by [36]:

$$d_{XRD} = \frac{0.93\,\lambda}{\Delta(2\theta)\cos(\theta)}$$

where d$_{XRD}$ is the mean size of the particles, λ is the X-ray wavelength (Co Kα, λ = 1.7927 Å) and Δ(2θ) is the line broadening at half the maximum intensity (FWHM) in radians. Using this model, the mean particle diameter d$_{XRD}$ for Fe$_{50}$Ni$_{50}$@CNT prepared by the first and second filling approaches equals to 8.9 ± 0.2 nm and 9.8 ± 0.2 nm, respectively. The discrepancy between the diameter measured by XRD and TEM can be explained by the possibility that some of the magnetic nanoparticles agglomerated, forming a larger polycrystalline structure, which appeared as a single larger particle in the TEM measurement. It is worth mentioning that the error bars in d$_{XRD}$ refer to the measurement error, while for d$_{TEM}$ it indicates the observed size distribution.

TGA is mainly used as a measure of the sample purity (i.e., the absence of outside particles which exhibit an increase in mass prior to the combustion of the CNTs), and for the determination of the filling material inside CNTs by performing backward calculations based on the mass of the TGA residue [37,38]. This measurement confirms the observation found by SEM that the main difference between the two filling approaches is the filling yield. That is, the samples prepared by the second filling approach are found to have a relatively higher filling yield in comparison with those prepared by the first filling approach. For some samples, the filling yield was about 5.0 ± 1 wt. % for solution-filled samples (Figure 5, black curve), whereas for samples prepared by the second approach, the filling yield was about 5.6 ± 1 wt. % (Figure 5, purple curve). The relatively lower filling yield of the solution-filled samples can be attributed to the occupancy of the inner volume of CNTs by water, which decreases the capacity of the inner CNTs cavity of filling, compared to the available filling volume in the second filling approach, in which only few drops of distilled water is required, solely to ensure good stirring (see Figure S2 in the Supplementary Materials).

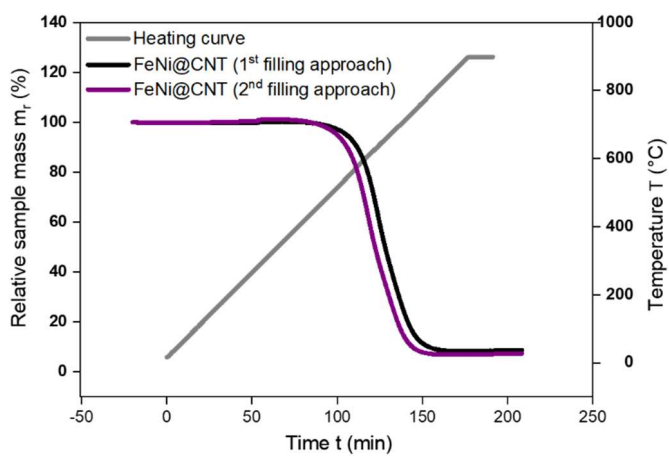

**Figure 5.** Relative sample mass loss for sample filled by the first (i.e., solution) approach (black) and the other by the second approach (purple) of $Fe_{50}Ni_{50}$@CNT during the combustion process of the nanocomposite, in which the CNTs mass start to decrease at T ~ 530 °C.

3.1.2. Magnetic Properties

The magnetic field dependence of the magnetization M(H) was measured for samples of $Fe_{50}Ni_{50}$@CNT prepared by both filling approaches, as shown in Figure 6a. Results for room temperature saturation magnetization $M_s$ obtained for $Fe_{50}Ni_{50}$@CNT are comparable with those obtained by Xu et al., who found the saturation magnetization at room temperature for equiatomic Fe-Ni alloy nanoparticles encapsulated in CNTs equals 141.66 emu/g [20]. In addition, saturation magnetization measured at 5 K was also compared with the data calculated from the Slater–Pauling curve at 0 K ($M_s$ = 165.7 emu/g) [39]. The magnetic data are summarized in Table 1. The data imply that the saturation magnetization measured at both temperatures agreed to a good extent with the reported data.

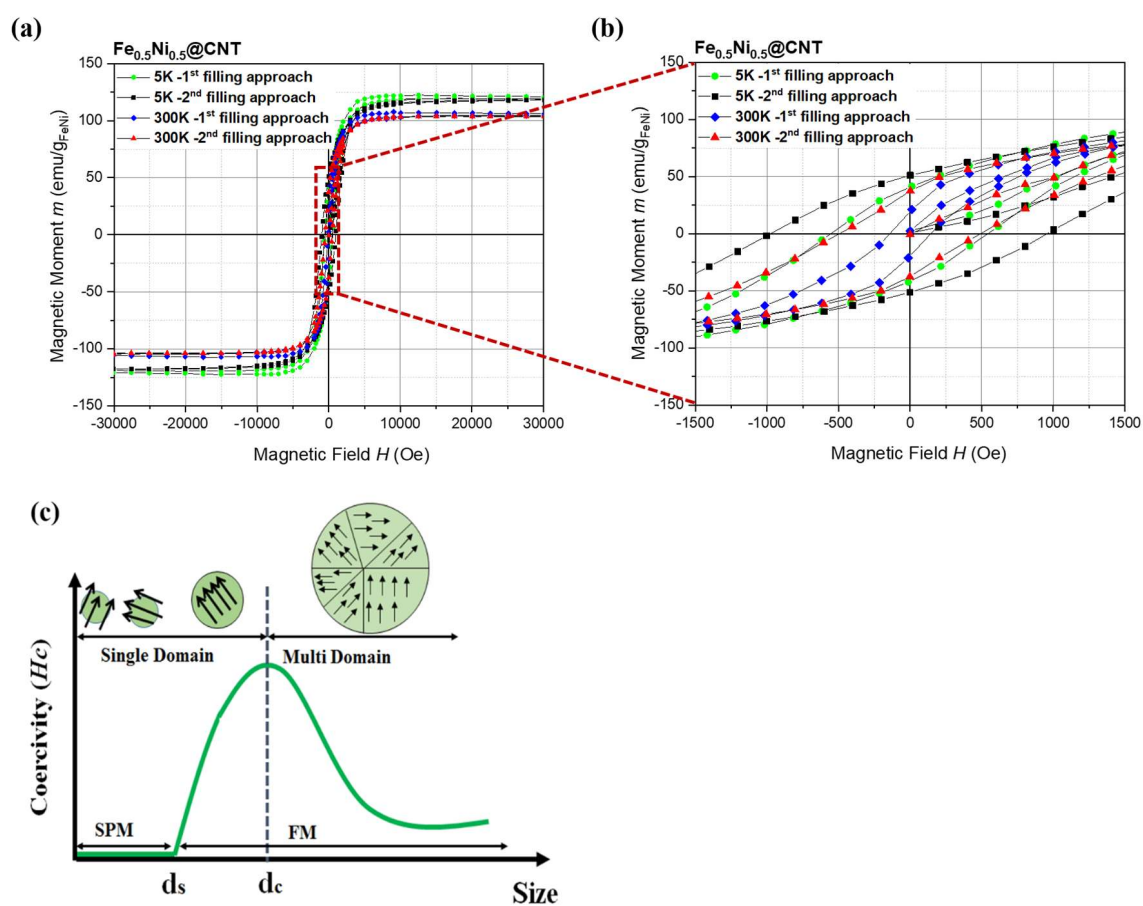

**Figure 6.** (**a**) Hysteresis curves for $Fe_{50}Ni_{50}$@CNT samples prepared by the first and second filling approaches. Data are normalized to the amount of magnetic material as obtained from the TGA measurements. (**b**) Enlarged view on the hysteresis curves show the enhancement in the coercive field for the nanoparticles. (**c**) Illustration showing the coercivity ($H_c$) as a function of particle diameter (D). Adapted from [40].

Further information on the magnetism of the materials is demonstrated by the observed difference in coercivity which is a size-dependent property already in the size range studied here. The data in Figure 6b demonstrate a great enhancement in the hardness of the magnetic nanoparticles prepared by both filling approaches compared to the reported bulk material.

A robust enhancement in the coercivity of the nanoparticles is observed compared to the bulk material of the same composition reported by Bozorth, who found the coercive field $H_c$ is less than 1 Oe for FeNi sample prepared by melting Armco iron and high-purity, commercial electrolytic nickel in a high-frequency induction furnace in air with a covering of borax [39]. The increase in the coercive field as the particle size decreases can be attributed to the size dependence of coercivity in the vicinity of the critical size of domain formation in nanoparticles. Briefly, in this size regime, coercivity of single domain (SD) magnets decrease upon size reduction while it increases in the multi-domain (MD) state (Figure 6c) [41,42]. On the other hand, our prepared nanoparticles also showed an enhancement in the coercivity compared to other nanoparticles prepared in the study by Xu et al. [20], in which $Fe_{50}Ni_{50}$ nanoparticles encapsulated inside CNTs via CVD process showed a coercivity of 68 Oe at room temperature, which is 2–7 times lower than the values obtained by our filling approaches. It is worth mentioning that nanoparticles prepared by our both filling approaches have mean particle diameter within the same range. However, nanoparticles prepared by the second approach show higher coercivity compared to those prepared by the first filling approach (Figure 6b). This can be attributed to the higher degree of filling (i.e., in the case of second filling approach), which produce particles with a pearl necklace-like appearance which constitute larger coercivity compared to the well-separated particles prepared by the first filling approach (Figure 1).

The symmetry in the shape of the hysteresis loops gives a good indication on the stability of the prepared samples against oxidation. As known, these particles are subject to oxidation when they are exposed to air unless they are shielded by the carbon shell. To be specific, the presence of oxide layers would imply the presence of an antiferromagnetic shell around the ferromagnetic cores, i.e. the material would evolve the exchange bias effect where nanoparticles cooled under magnetic field show a significant shift between the coercive field values at the positive ($H_{c+}$) and the negative ($H_{c-}$) sides [24,41,43]. For the prepared samples, and within the experimental errors, equal values of $H_{c+}$ and $H_{c-}$ have been found (Table 1), a further indication for the protective shells of CNTs.

Table 1. Physical properties of the magnetic nanoparticles of $Fe_{50}Ni_{50}$@CNT.

| Parameter | Filling Approach | |
| --- | --- | --- |
| | First Approach | Second Approach |
| $d_{TEM}$ (nm) | 19 ± 10 | 17 ± 4 |
| $d_{XRD}$ (nm) | 8.9 ± 0.2 | 9.8 ± 0.2 |
| TGA (wt. %) | 5.0 ± 1 | 5.6 ± 1 |
| $M_s$ (emu/$g_{FeNi}$) (300K) | 108 ± 22 | 104 ± 19 |
| $M_s$ (emu/$g_{FeNi}$) (5K) | 123 ± 25 | 118 ± 21 |
| $H_c$ (Oe) (300K) | 143 ± 2 | 492 ± 3 |
| $H_c$ (Oe) (5K) | 548 ± 7 | 967 ± 12 |

*3.2. $Fe_{67}Ni_{33}$@CNT*

Based on the results obtained so far, the second filling approach has the advantage that, in comparison to the first one, a relatively higher degree of filling could be obtained. As a result, this approach (i.e., second filling approach) has been adopted for the preparation of $Fe_{67}Ni_{33}$@CNT samples. Therefore, all data presented in this section are related to $Fe_{67}Ni_{33}$@CNT samples prepared by the second filling approach.

3.2.1. Morphology and Structure

The filling behavior of $Fe_{50}Ni_{50}$@CNT has been also observed for the as-prepared sample of $Fe_{67}Ni_{33}$@CNT (Figure 7a), in which the filling particles are distributed along the inner cavity of the CNTs. However, it is important to emphasize the effect of annealing on the growth of the $Fe_{67}Ni_{33}$@CNT nanoparticles. For the as-prepared sample, different morphologies and particle sizes for the filling materials were observed (small and large spheres, particle chains as indicated by arrows in Figure 7a), whereas, after an additional heat treatment step (annealing at 500 °C for 12 h), a significant growth in the particles size was observed (Figure 7b). This observation was confirmed by statistical measurements of the particles aspect ratios and their diameters with respect to the diameter of the inner walls for the CNTs. We attribute the observation of a pronounced increase in particle size for a prolonged heat treatment (i.e., annealing for 12 h) to the availability of a significantly higher amount of thermal energy, which increase the probability of particle merging, which in turn leads to particles growth.

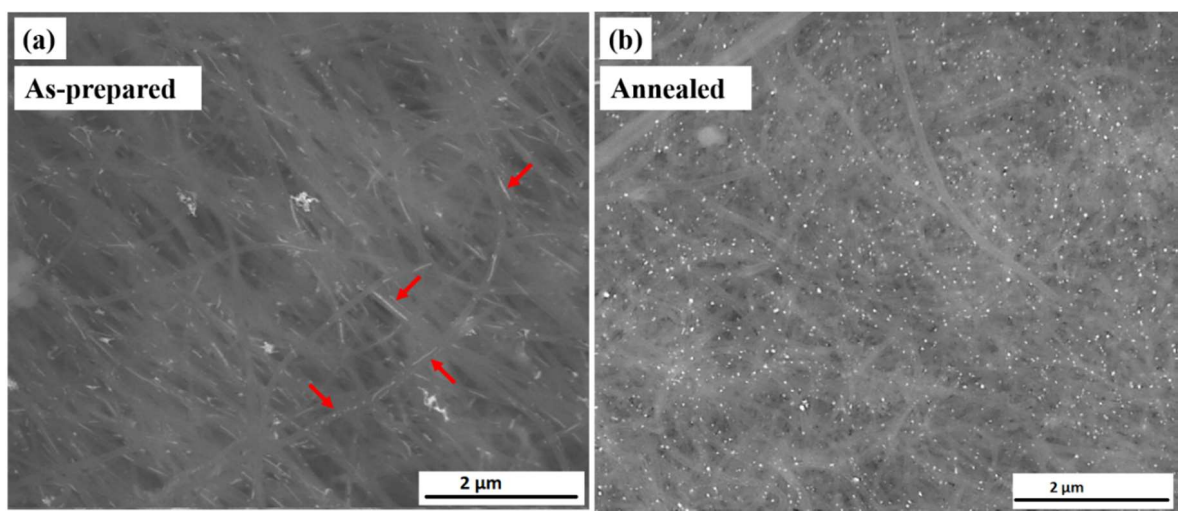

**Figure 7.** SEM overview images in BSE contrast of the (**a**) as-prepared and (**b**) annealed samples of Fe$_{67}$Ni$_{33}$@CNT prepared by the second filling approach.

TEM measurements were performed for the as-prepared and annealed samples of Fe$_{67}$Ni$_{33}$@CNT. Figure 8a is an example of an as-prepared sample of Fe$_{67}$Ni$_{33}$@CNT, in which most of the particles are located within the hollow cavity of the CNTs and exhibit a broad variety of sizes and morphologies. After further heat treatments (Figure 8b), homogeneous morphologies and particles size were observed. It is obvious that the annealing has an impact on the growth of particles size to a diameter close to the inner CNTs diameter.

TEM-based nanobeam electron diffraction carried out for several individual nanoparticles for the annealed sample revealed that the filling particles are single crystalline. Similar to the FeNi alloys containing 30 at. % Ni or higher, the annealed sample showed the typical fcc structures of γ-FeNi indicated by the reflections correspond to the 111 (0.207 nm), 200 (0.179 nm), 220 (0.128 nm), 311 (0.109 nm) and 222 (0.104 nm) lattice planes (Figure 8c) [17]. It is worth mentioning that the bcc structure of α-FeNi has also been observed for a number of selected particles. This was indicated by the reflections corresponding to the 110 (0.203 nm), 200 (0.145 nm), 220 (0.103 nm) and 310 (0.092 nm) lattice planes (Figure 8d). In other words, both fcc and bcc structures coexist in the Fe-rich samples. The results were also confirmed by XRD data shown below. We could not obtain nanobeam electron diffraction patterns for the as-prepared samples, this can be attributed to the inhomogeneity of the particles, lack of crystallinity, and the presence of different phases inside CNTs. This finding underlines that a further annealing step is essential to obtain well-defined intermetallic nanoparticles.

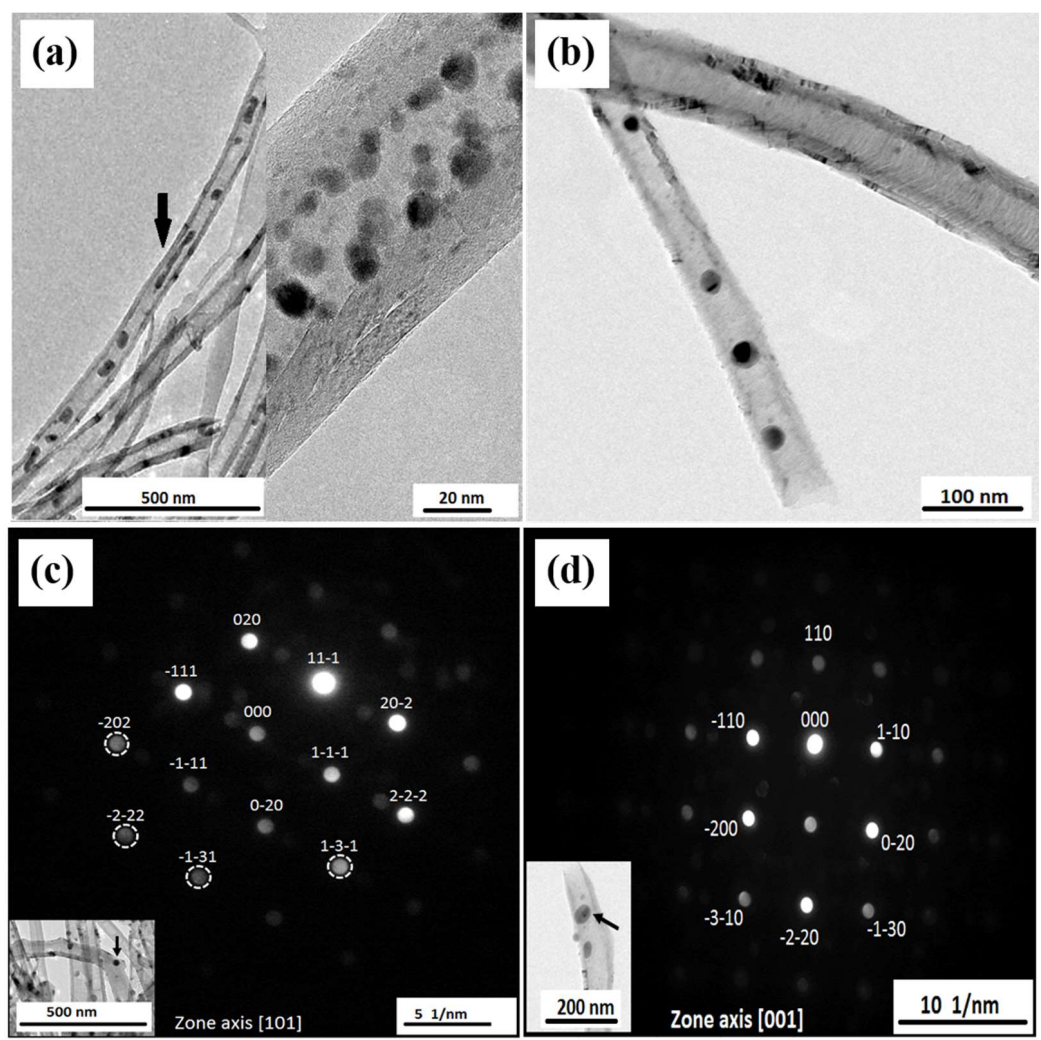

**Figure 8.** TEM bright field images for the (**a**) as-prepared and (**b**) annealed samples of Fe$_{67}$Ni$_{33}$@CNT prepared by the second filling approach. (**c**,**d**) Nanobeam electron diffraction patterns of the annealed Fe$_{67}$Ni$_{33}$ nanoparticles with the corresponding TEM images as insets.

The distribution of the particles diameter was investigated for the as-prepared and annealed samples. As-prepared samples had a mean diameter of $d_{TEM}$ = 19 ± 8 nm (see Figure S3 in the Supplementary Materials), whereas the annealed samples, a slight change in the mean diameter was observed, in which the mean diameter of $d_{TEM}$ = 28 ± 9 nm was found. Comparing these values with the mean diameter of the CNTs ($d_{CNT}$ = 42 ± 14 nm), one can conclude that the confinement of the magnetic nanoparticles to the inner diameter of the CNTs allows control of the particle size, and to a large extent prevent particles agglomeration in both samples.

Similar to Fe$_{50}$Ni$_{50}$@CNT, the spherical geometry of the Fe$_{67}$Ni$_{33}$ nanoparticles inside CNTs was confirmed by aspect-ratio studies.

In an as-prepared sample (with nearly 87 investigated filling particles), 55% of the particles had aspect ratios of about 1–1.2, whereas 31% of the particles had ratios in the range of 1.2–1.5. The remaining 14% exhibited ratios larger than 1.5. However, in an annealed sample (with nearly 139 investigated filling particles), 72% of the particles had aspect ratios of about 1–1.2, whereas 25% had aspect ratios in the range of 1.2–1.5, and only 3% exhibited ratios larger than 1.5. This means that the majority of the particles became spherical, and, hence, the morphology became more homogenous after further heat treatment. We attribute the homogeneity in the morphology for the annealed samples to the non-wetting behavior between FeNi alloy nanoparticles and carbon nanotubes [26], in which the contact area between the nanoparticles and the tubes tend to minimize. Hence, when annealing takes place, the filling particles tend to agglomerate and form spherical particles. A schematic representation of the geometry of the filling particles based on the aspect ratio values is already shown in Figure 3c.

The expected stoichiometry of 2:1 for the binary alloys was confirmed by EDX measurements in a similar way to the equimolar FeNi binary alloys. Quantitative analysis indicates that the average atomic percentage of Fe is 67.0 ± 3.0 at. % and for Ni is 33.0 ± 3.0 at. % in $Fe_{67}Ni_{33}$@CNT samples.

XRD measurements were performed for the as-prepared and annealed samples of $Fe_{67}Ni_{33}$@CNT (Figure 9). FeNi alloys containing 30 at. % Ni or higher, showed the typical fcc-crystal structures of $\gamma$-FeNi, in which the reflections at 2θ~51° and 60° correspond to the lattice planes 111 and 200, respectively, of the fcc structure of FeNi with space group Fm-3m (225, cubic, PDF No. 04-004-8844) [17]. However, unlike equimolar $Fe_{50}Ni_{50}$@CNT samples, both fcc and bcc phases coexist in the Fe-rich samples. These results were also observed by Wu et al. [3] and Sumiyama et al. [44] for samples containing Fe content higher than 50 at. %. As a result, the reflections at 2θ ~ 52° and 77° correspond to the lattice planes 110 and 200, respectively, of the bcc structure of $\alpha$-FeNi with space group Im-3m (229, cubic, PDF No. 04-004-5110). No reflections corresponding to oxides or carbides were detected. It is worth to mention, that the bcc structure could not be observed in the as-prepared sample. This is an indication that the Fe-rich compound of $Fe_{67}Ni_{33}$@CNT was not formed. However, after an additional annealing step, an additional reflection of 110 could be observed, an indication of the formation of the $Fe_{67}Ni_{33}$@CNT compound.

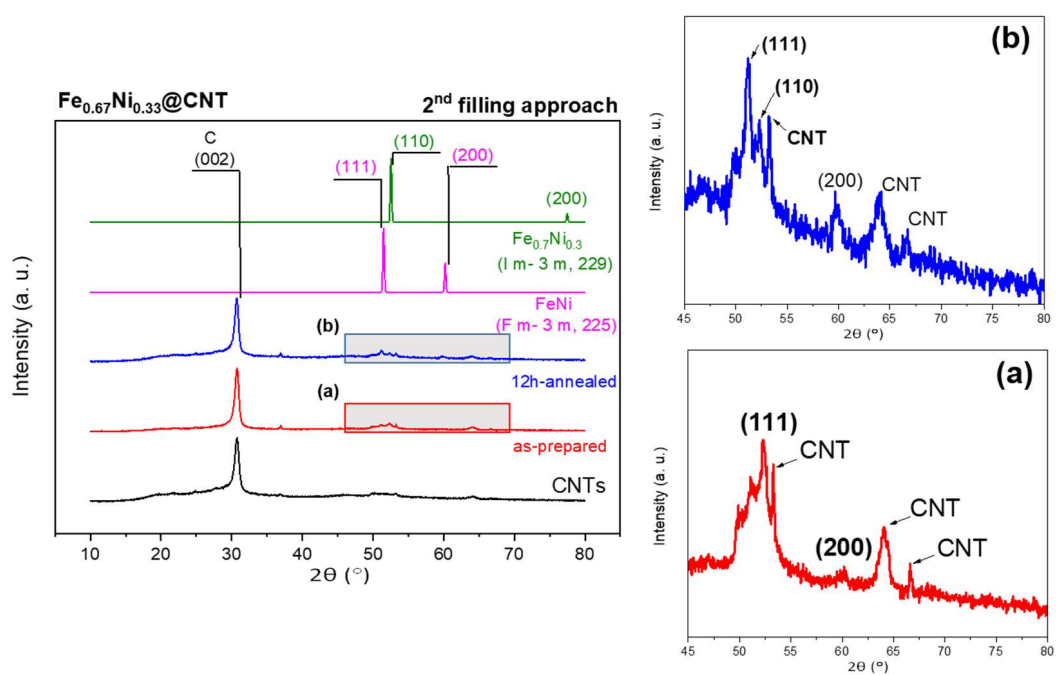

**Figure 9.** XRD diffraction pattern for the as-prepared and annealed samples of $Fe_{67}Ni_{33}$@CNT (*left-hand side*). Enlarged views (marked by rectangles) for the main reflections for the (**a**) as-prepared and (**b**) annealed samples, are shown on the right-hand side.

The mean particle diameter for the as-prepared and annealed samples was also calculated using Scherrer's equation. Using this model, the mean particle diameter $d_{XRD}$ for the as-prepared samples equals to 3.0 ± 0.3 nm, whereas after an additional annealing step, a slight increase in the particle sizes was observed, in which $d_{XRD}$ equals to 9.0 ± 0.4 nm. As mentioned earlier, the discrepancy between the diameter measured by XRD and TEM can be explained by the possibility that some of the magnetic nanoparticles agglomerated, forming a larger polycrystalline structure, which appeared as a single larger particle in the TEM measurement. The error bars in $d_{XRD}$ refer to the measurement error while for $d_{TEM}$ they indicate the observed size distribution.

TGA measurements showed a filling yield of about 2.70 ± 1 wt. % for the annealed Fe$_{67}$Ni$_{33}$@CNT samples prepared by the second filling approach (see Figure S4 in the Supplementary Materials).

3.2.2. Magnetic Properties

The magnetic field dependence of the magnetization M(H) has been measured for the as-prepared and annealed samples of Fe$_{67}$Ni$_{33}$@CNT as shown in Figure 10a. Saturation magnetization M$_s$ measured at 5K for the annealed samples correlates well with M$_s$ calculated from Slater-Pauling curve for FeNi alloys of the same stoichiometry (~186.0 emu/g) [39]. Further, the magnetization data gives evidence for the importance of the annealing step, since the as-prepared samples exhibit M$_s$ significantly lower than the reported data for the bulk material of the same stoichiometry [32]. This may be attributed to the lower crystallinity of the MNPs for the as-prepared samples compared to the annealed samples, and the formation of a bulk-like ferromagnetic core and a shell composed of disordered moments [21,45].

The fact that the as-prepared nanoparticles have significantly larger coercivities compared to the annealed nanoparticles can be straightforwardly attributed to the small size of the MNPs stabilizing the SD state (Figure 10b). The larger post-annealed MNPs exhibit smaller coercive fields can hence be associated with the MD state realized in such large particles (Figure 6c).

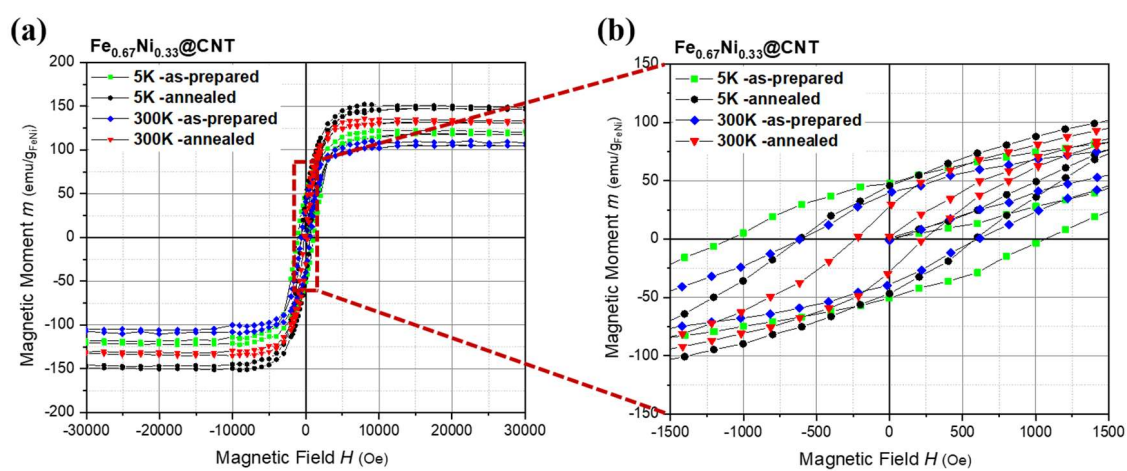

**Figure 10**. (**a**) Hysteresis curves for the as-prepared and annealed samples of Fe$_{67}$Ni$_{33}$@CNT prepared by second filling approach. Data are normalized to the amount of magnetic material as obtained from the TGA measurements. (**b**) Enlarged view on the hysteresis curves.

Similar to Fe$_{50}$Ni$_{50}$@CNT hysteresis curves, and as emphasis on the protective shells of CNTs against nanoparticles oxidation, the symmetry in the shape of the hysteresis loops gives a good indication on the stability of the prepared samples against oxidation. For the as-prepared and annealed samples, and within the experimental errors, equal values of H$_{c+}$ and H$_{c-}$ were found.

**Table 2** summarizes all the data obtained for Fe$_{67}$Ni$_{33}$@CNT.

**Table 2.** Physical properties of the magnetic nanoparticles of $Fe_{67}Ni_{33}$@CNT.

| Parameter | Second Approach | |
|---|---|---|
| | As-Prepared | Annealed |
| $d_{TEM}$ (nm) | 19 ± 8 | 28 ± 9 |
| $d_{XRD}$ (nm) | 3.0 ± 0.3 | 9.0 ± 0.4 |
| TGA (wt. %) | 2.7 ± 1 | |
| $M_s$ (emu/$g_{FeNi}$) (300K) | 110 ± 40 | 136 ± 50 |
| $M_s$ (emu/$g_{FeNi}$) (5K) | 123 ± 47 | 152 ± 56 |
| $H_c$ (Oe) (300K) | 608 ± 1 | 234 ± 1 |
| $H_c$ (Oe) (5K) | 1080 ± 16 | 601 ± 17 |

## 4. Conclusions

In this study, two types of FeNi alloys nanoparticles were successfully encapsulated within the hollow cavity of CNTs using two facile routes. Our study demonstrates that both wet chemical methods offer simple ways to fill CNTs with these nanoparticles in a well-defined manner without the need of vigorous conditions. Depending on the chosen procedure, we were able to influence the filing yield, size, magnetic moment, coercivity, and the appearance of the sample in terms of particles inside CNTs to those on the outer surface. Tuning several parameters, such as mass of CNTs, precursor concentration, type and volume of washing agents, reduction/annealing time and temperature, etc., resulted in CNTs filled with FeNi nanoparticles in a well-defined manner. These facile approaches can be extended for the filling of CNTs with many other kinds of magnetic nanoparticles. However, it should be taken into consideration that, for some stoichiometries, additional heating steps are required in order to obtain the right phase.

The crystallinity of the Fe-Ni nanoparticles in both types of binary alloys was verified by powder XRD and nanobeam electron diffraction. No indication of oxides or carbides phases were detected, which means that the synthesis approaches guarantee CNTs as protective shells for the MNPs. The additional annealing step for $Fe_{67}Ni_{33}$@CNT sample is mandatory for pure phase particles. The magnetic properties for the nanoparticles show preservation in the saturation magnetization at the nanoscale, whereas the hardness of the nanoparticles can be manipulated depending on the filling approach. The produced materials can be considered as excellent candidates for different applications. For example, industrial applications which require high coercivities (such as magnetic storage devices), and medical applications (where they are considered as good thermos-seed candidates for cancer therapy) could benefit from these enhanced materials.

**Supplementary Materials**: The following are available online at www.mdpi.com/. Figure S1: SEM-EDX quantitative measurement for the Fe:Ni ratio over a selected area for a sample of $Fe_{50}Ni_{50}$@CNT, in which the ratio roughly corresponds to 1:1, Figure S2: Relative sample mass loss for the pristine CNTs (Pyrograf) during the combustion process, in which 100 % of the CNTs mass has been lost, Figure S3: Histograms representing the size distribution of the inner diameter (nm) of CNTs and particles diameters for the (**a**) as-prepared and (**b**) annealed samples of Fe$_{67}$Ni$_{33}$@CNT, Figure S4: Relative sample mass loss for Fe$_{67}$Ni$_{33}$@CNT sample filled by the second approach during the combustion process of the nanocomposite, in which the CNTs mass start to decrease at T ~ 530 °C.

**Author Contributions:** Conceptualization, R. G.; Methodology, R. G.; Formal Analysis, R. G.; Writing-Original Draft Preparation, C. D.; TEM and HRTEM measurements, D. W. and A. L.; TEM tomography, S. H.; Writing-Review & Editing and group leader, B. B., and M. M.; Supervision.

**Funding:** This research received no external funding.

**Acknowledgments:** The authors thank G. Kreutzer for additional TEM measurements. R.G. acknowledges the German Academic Exchange Service (DAAD) for funding. A.L. and D.W. acknowledge funding from the European Research Council via the ERC-2016-STG starting grant ATOM. The publication of this article was funded by the Open Access Fund of the Leibniz Association.

**Conflicts of Interest:** The authors declare no conflict of interest.

**References**

1. Nalwa, H.S. *Handbook of nanostructured materials and nanotechnology*; Academic Press: Cambridge, Massachusetts, USA, 1999.
2. McHenry, M.E.; Willard, M.A.; Laughlin, D.E. Amorphous and nanocrystalline materials for applications as soft magnets. *Prog. Mater. Sci.* **1999,** *44*, 291–433.
3. Wu, H.-Q.; Cao, Y.-J.; Yuan, P.-S.; Xu, H.-Y.; Wei, X.-W. Controlled synthesis, structure and magnetic properties of Fe$_{1-x}$Ni$_x$ alloy nanoparticles attached on carbon nanotubes. *Chem. Phys. Lett.* **2005,** *406*, 148–153.
4. Steigerwalt, E.S.; Deluga, G.A.; Lukehart, C. Pt-Ru/carbon fiber nanocomposites: Synthesis, characterization, and performance as anode catalysts of direct methanol fuel cells. A search for exceptional performance. *J. Phys. Chem. B.* **2002**, *106*, 760–766.
5. Wu, H.-Y.; Zhao, Y.; Jiao, Q.-Z. Nanotube arrays of Zn/Co/Fe composite oxides assembled in porous anodic alumina and their magnetic properties. *J. Alloy. Compd.* **2009**, *487*, 591–594.
6. Winkler, A.; Mühl, T.; Menzel, S.; Kozhuharova-Koseva, R.; Hampel, S.; Leonhardt, A.; Büchner, B. Magnetic force microscopy sensors using iron-filled carbon nanotubes. *J. appl. phys.* **2006**, *99*, 104905.
7. Mönch, I.; Meye, A.; Leonhardt, A.; Krämer, K.; Kozhuharova, R.; Gemming, T.; Wirth, M.P.; Büchner, B. Ferromagnetic filled carbon nanotubes and nanoparticles: Synthesis and lipid-mediated delivery into human tumor cells. *J. Magn. Magn. Mater.* **2005**, *290*, 276–278.
8. Pankhurst, Q.A.; Connolly, J.; Jones, S.; Dobson, J. Applications of magnetic nanoparticles in biomedicine. *J. Phys. D: Appl. Phys.* **2003**, *36*, R167.
9. Klingeler, R.; Hampel, S.; Buchner, B. Carbon nanotube based biomedical agents for heating, temperature sensoring and drug delivery. *Int. J. Hyperth.* **2008**, *24*, 496–505.
10. Gonzalez, E.; Jasen, P.; Gonzalez, G.; Moro, L.; Juan, A. Hydrogen and carbon interaction in a FeNi alloy with a vacancy. *Phys. Status. Solidi. B.* **2009**, *246*, 1275–1285.
11. Gheisari, K.; Javadpour, S.; Oh, J.; Ghaffari, M. The effect of milling speed on the structural properties of mechanically alloyed Fe–45% Ni powders. *J. Alloy. Compd.* **2009**, *472*, 416–420.
12. McNerny, K.L.; Kim, Y.; Laughlin, D.E.; McHenry, M.E. Chemical synthesis of monodisperse γ-Fe–Ni magnetic nanoparticles with tunable Curie temperatures for self-regulated hyperthermia. *J. Appl. Phys.* **2010**, *107*, 09A312.
13. Zheng, X.; Deng, J.; Wang, N.; Deng, D.; Zhang, W.H.; Bao, X.; Li, C. Podlike N-doped carbon nanotubes encapsulating FeNi alloy nanoparticles: High-performance counter electrode materials for dye-sensitized solar cells. *Angew. Chem.* **2014**, *53*, 7023–7027.
14. Arnold, H.; Elmen, G. Permalloy, a new magnetic material of very high permeability. *Bell. Labs. Tech. J.* **1923**, *2*, 101–111.
15. Lv, R.; Kang, F.; Cai, D.; Wang, C.; Gu, J.; Wang, K.; Wu, D. Long continuous FeNi nanowires inside carbon nanotubes: Synthesis, property and application. *J. Phys. Chem. Solid.* **2008**, *69*, 1213–1217.

16. Guillaume, C.É. Recherches sur les aciers au nickel. Dilatations aux temperatures elevees; resistance electrique. *CR. Acad. Sci*. **1897**, *125*, 18.
17. Grobert, N.; Mayne, M.; Walton, D.R.M.; Kroto, H.W.; Terrones, M.; Kamalakaran, R.; Seeger, T.; Rühle, M.; Terrones, H.; Sloan, J.; Dunin-Borkowski, R.E.; Hutchison, J.L. Alloy nanowires: Invar inside carbon nanotubes. *Chem. Commun*. **2001**, *5*, 471–472.
18. Bantz, C.; Koshkina, O.; Lang, T.; Galla, H.-J.; Kirkpatrick, C.J.; Stauber, R.H.; Maskos, M. The surface properties of nanoparticles determine the agglomeration state and the size of the particles under physiological conditions. *Beilstein. J. Nanotech*. **2014**, *5*, 1774–1786.
19. Issa, B.; Obaidat, I.M.; Albiss, B.A.; Haik, Y. Magnetic nanoparticles: Surface effects and properties related to biomedicine applications. *Int. J. Mol. Sci*. **2013**, *14*, 21266–21305.
20. Xu, M.H.; Zhong, W.; Qi, X. S.; Au, C. T.; Deng, Y.; Du, Y. W. Highly stable Fe–Ni alloy nanoparticles encapsulated in carbon nanotubes: Synthesis, structure and magnetic properties. *J. Alloy. Compd*. **2010**, *495*, 200–204.
21. Ghunaim, R.; Eckert, V.; Scholz, M.; Gellesch, M.; Wurmehl, S.; Damm, C.; Buechner, B.; Mertig, M.; Hampel, S. Carbon nanotube-assisted synthesis of ferromagnetic Heusler nanoparticles of $Fe_3Ga$ (Nano-Galfenol). *J. Mater. Chem. C*. **2018**, *6*, 1255–1263.
22. Costa, P.M.; Gautam, U.K.; Bando, Y.; Golberg, D. Comparative study of the stability of sulfide materials encapsulated in and expelled from multi-walled carbon nanotube capsules. *Carbon* **2011**, *49*, 342–346.
23. Lipert, K.; Bahr, S.; Wolny, F.; Atkinson, P.; Weißker, U.; Mühl, T.; Schmidt, O.; Büchner, B.; Klingeler, R. An individual iron nanowire-filled carbon nanotube probed by micro-Hall magnetometry. *Appl. Phys. Lett.* **2010**, *97*, 212503.
24. Ibrahim, E.M.M.; Hampel, S.; Wolter, A.U.B.; Kath, M.; El-Gendy, A.A.; Klingeler, R.; Täschner, C.; Khavrus, V.O.; Gemming, T.; Leonhardt, A.; Büchner, B. Superparamagnetic FeCo and FeNi Nanocomposites Dispersed in Submicrometer-Sized C. Spheres. *J. Phys. Chem. C*. **2012**, *116*, 22509–22517.
25. Sloan, J.; Wright, D.M.; Woo, H.G.; Bailey, S.; Brown, G.; York, A.P.E.; Coleman, K.S.; Hutchison, J.L.; Green, M.L.H. Capillarity and silver nanowire formation observed in single walled carbon nanotubes. *Chem. Commun*. **1999**, *8*, 699–700.
26. Dujardin, E.; Ebbesen, T.; Hiura, H.; Tanigaki, K. Capillarity and wetting of carbon nanotubes. *Science* **1994**, *265*, 1850–1852.
27. Ajayan, P.M. Capillarity-induced filling of carbon nanotubes. *Nature* **1993**, *361*, 333–334.
28. Wu, H.-Q.; Wei, X.-W.; Shao, M.-W.; Gu, J.-S.; Qu, M.-Z. Preparation of Fe–Ni alloy nanoparticles inside carbon nanotubes via wet chemistry. *J. Mater. Chem.* **2002**, *12*, 1919–1921.
29. Pyrograf Products. Available online: www.pyrografproducts.com (accessed on 5 July 2018).
30. Arlt, M.; Haase, D.; Hampel, S.; Oswald, S.; Bachmatiuk, A.; Klingeler, R.; Schulze, R.; Ritschel, M.; Leonhardt, A.; Fuessel, S.; Buchner, B.; Kraemer, K.; Wirth, M. P. Delivery of carboplatin by carbon-based nanocontainers mediates increased cancer cell death. *Nanotechnology* **2010**, *21*, 335101.
31. Tessonnier, J.-P.; Rosenthal, D.; Hansen, T. W.; Hess, C.; Schuster, M. E.; Blume, R.; Girgsdies, F.; Pfänder, N.; Timpe, O.; Su, D. S. Analysis of the structure and chemical properties of some commercial carbon nanostructures. *Carbon* **2009**, *47*, 1779–1798.
32. Ghunaim, R.; Scholz, M.; Damm, C.; Rellinghaus, B.; Klingeler, R.; Büchner, B.; Mertig, M.; Hampel, S. Single-crystalline FeCo nanoparticle-filled carbon nanotubes: Synthesis, structural characterization and magnetic properties. *Beilstein J. Nanotech*. **2018**, *9*, 1024–1034.
33. Tsang, S.; Chen, Y.; Harris, P.; Green, M. A simple chemical method of opening and filling carbon nanotubes. *Nature* **1994**, *372*, 159–162.
34. Gellesch, M. Statistical study of the effect of annealing treatments on assemblies of intermetallic magnetic nanoparticles related to the Heusler compound Co 2 FeGa. Ph.D. Thesis, Technische Universität Dresden, Dresden, Saxony, Germany, 2016.
35. Costa, P.M.; Sloan, J.; Rutherford, T.; Green, M. L. Encapsulation of $Re_xO_y$ clusters within single-walled carbon nanotubes and their in tubulo reduction and sintering to Re metal. *Chem. Mater*. **2005**, *17*, 6579–6582.
36. Patterson, A. The Scherrer formula for X-ray particle size determination. *Phys. Rev*. **1939**, *56*, 978.
37. Haft, M.; Grönke, M.; Gellesch, M.; Wurmehl, S.; Büchner, B.; Mertig, M.; Hampel, S. Tailored nanoparticles and wires of Sn, Ge and Pb inside carbon nanotubes. *Carbon* **2016,** *101*, 352–360.

38. Gellesch, M.; Dimitrakopoulou, M.; Scholz, M.; Blum, C.G.F.; Schulze, M.; van den Brink, J.; Hampel, S.; Wurmehl, S.; Büchner, B. Facile Nanotube-Assisted Synthesis of Ternary Intermetallic Nanocrystals of the Ferromagnetic Heusler Phase Co2FeGa. *Cryst. Growth. Des*. **2013,** *13*, 2707–2710.
39. Bozorth, R. M. Ferromagnetism. *Wiley-VCH* **1993**, *1*, 992.
40. Akbarzadeh, A.; Samiei, M.; Davaran, S. Magnetic nanoparticles: Preparation, physical properties, and applications in biomedicine. *Nanoscale. Res. Lett*. **2012**, *7*, 144.
41. Lee, J.S.; Myung, Cha J.; Young Yoon, H.; Lee, J. K.; Kim, Y. K. Magnetic multi-granule nanoclusters: A model system that exhibits universal size effect of magnetic coercivity. *Sci. Rep*. **2015**, *5*, 12135.
42. Kneller, E.F.; Luborsky, F.E. Particle Size Dependence of Coercivity and Remanence of Single-Domain Particles. *J. Appl. Phys*. **1963**, *34*, 656–658.
43. Nogués, J.; Schuller, I.K. Exchange bias. *J. Magn. Magn. Mater*. **1999**, *192*, 203–232.
44. Sumiyama, K.; Kadono, M.; Nakamura, Y. Metastable bcc phase in sputtered Fe–Ni alloys. *Trans. Jpn. Inst. Met*. **1983**, *24*, 190–194.
45. Ibrahim, E.; Abdel-Rahman, L.H.; Abu-Dief, A.M.; Elshafaie, A.; Hamdan, S.K.; Ahmed, A. Electric, thermoelectric and magnetic characterization of γ-$Fe_2O_3$ and $Co_3O_4$ nanoparticles synthesized by facile thermal decomposition of metal-Schiff base complexes. *Mater. Res. Bulletin*. **2018**, *99*, 103–108.

.
.
.